\documentclass[twoside,showpacs,superscriptaddress,twocolumn,floatfix,a4paper,aps,pra]{revtex4}

\usepackage{color}
\usepackage{graphicx}
\usepackage[utf8]{inputenc}

\usepackage{amssymb}
\usepackage{hyperref}
\usepackage{siunitx}

\begin{document}

\title{Scheme for a linear-optical controlled-phase gate with programmable phase shift}

\author{Karel Lemr} \email{k.lemr@upol.cz}
\affiliation{RCPTM, Joint Laboratory of Optics of Palacký University and Institute of Physics of Academy of Sciences of the Czech Republic, 17. listopadu 12, 771 46 Olomouc, Czech Republic}

\author{Karol Bartkiewicz} 
\affiliation{Faculty of Physics, Adam Mickiewicz University,
PL-61-614 Pozna\'n, Poland}
\affiliation{RCPTM, Joint Laboratory of Optics of Palacký University and Institute of Physics of Academy of Sciences of the Czech Republic, 17. listopadu 12, 771 46 Olomouc, Czech Republic}

\author{Antonín Černoch} 
\affiliation{RCPTM, Joint Laboratory of Optics of Palacký University and Institute of Physics of Academy of Sciences of the Czech Republic, 17. listopadu 12, 771 46 Olomouc, Czech Republic}

\date{\today}

\begin{abstract}
We present a linear-optical scheme for a controlled-phase gate with tunable phase shift programmed by a qubit state. In contrast to all previous tunable controlled-phase gates, the phase shift is not hard-coded into the optical setup, but can be tuned to any value from 0 to $\pi$ by the state of a so-called program qubit. Our setup is feasible with current level of technology using only linear-optical components. We provide an experimental feasibility study to assess the gate's implementability. We also discuss options for increasing the success probability up to 1/12 which approaches the success probability of a optimal non-programmable tunable controlled-phase gate.
\end{abstract}


\maketitle

\section{Introduction}
\label{Introduction}
Quantum computing is a promising approach allowing, in principle, considerably increasing computing efficiency \cite{Nielsen_QCQI,Zeilinger_QIP}. It has been demonstrated that any quantum circuit can be decomposed into a set of standard single and two-qubit gates \cite{Barenco95universal}. While the single-qubit gates represent just single qubit rotations, the two-qubit gates make the qubits interact and thus process the information. A prominent example of such a two-qubit gate is the controlled-phase gate (or its close relative the controlled-NOT gate) \cite{Shende04optim}.

The controlled-phase gate performs the following transformation on the target and control qubit states:
\begin{eqnarray}
\label{eq:cphase}
|00\rangle	&\rightarrow&|00\rangle\nonumber,\\
|01\rangle	&\rightarrow&|01\rangle\nonumber,\\
|10\rangle	&\rightarrow&|10\rangle\nonumber,\\
|11\rangle	&\rightarrow&\mathrm{e}^{i\varphi}|11\rangle,
\end{eqnarray}
where 0 and 1 in the brackets stand for logical states of the target and control qubits respectively. The parameter $\varphi$ then denotes the introduced phase shift. There have been a number of experimental implementations of the controlled-phase gate achieved on various physical platforms including nuclear magnetic resonance \cite{Cory97}, trapped ions \cite{Schmidt-Kaler03} or superconducting qubits \cite{Plantenberg07}. On the platform of linear optics, this gate has been implemented using various schemes \cite{Ralph02,OBrien03,Kiesel05} (for review papers see also \cite{Kok07,Bartkowiak10}). All these implementations however only considered phase shift $\varphi = \pi$ also known as the controlled-sign transformation.

\begin{figure}
\includegraphics[scale=1]{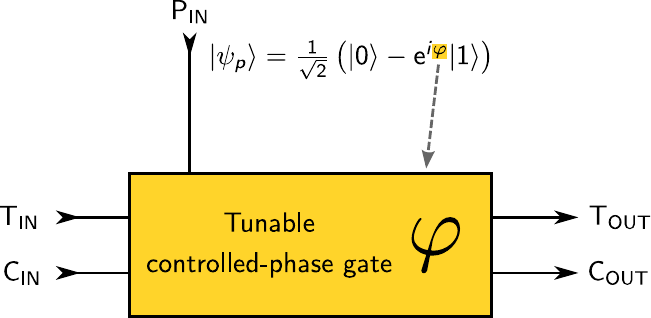}
\caption{\label{fig:concept} Conceptual scheme of a programmable c-phase gate. ``T", ``C" and ``P" denote target, control and program ports respectively. The phase shift $\varphi$ encoded into the state of the program qubit translates into the phase shift introduced by the gate according to the Eq. (\ref{eq:cphase}).}
\end{figure}
Operating the controlled-phase gate at phase shifts other then $\pi$ has been investigated for the first time in a seminal paper by Lanyon {\it et al.} from 2009 \cite{Lanyon09cpg}. In order to achieve divers phase shifts, the authors increased the Hilbert space by introducing axillary modes. Their implementation however does not have optimal success probability. In 2010, Konrad Kieling and his colleagues proposed a scheme for optimal linear-optical c-phase gate with tunable phase shift \cite{Kieling10cpg}. In 2011, this scheme has been experimentally implemented and tested in our laboratory \cite{Lemr11cpg}. Both our \cite{Lemr11cpg} and the Lanyon {\it et al.} \cite{Lanyon09cpg} scheme have the phase-shift hard-coded by the specific setting of various optical elements. This fact limits the gates in their adaptability and use in multi-purpose quantum circuits.

In order to make quantum circuits more versatile, researchers have proposed the so-called programmable gates \cite{proggates}. Instead of hard-coding the transformations into the experimental setup, these gates have their properties programmed by quantum state of the so-called program qubit. While it would be necessary to use infinite amount of classical information, to precisely set a real-valued parameter of a quantum gate (or transformation), one qubit of quantum information suffices. Such qubit can also be transmitted over a quantum channel thus allowing for remote programming of a quantum gate similar to classical software distribution over computer networks. The quantum transformation in question is basically teleported to its user.

As a proof-of-principle, Mičuda {\it et al.} have constructed a programmable phase gate \cite{Micuda08}. This gate introduces a programmable phase shift between logical states $|0\rangle$ and $|1\rangle$ of a signal qubit and thus it achieves programmable single-qubit rotation along one axis. The success probability of this scheme has been recently improved to the theoretical limit of $1/2$ using feed-forward \cite{Mikova12}.

In this paper, we propose a linear-optical scheme for a tunable c-phase gate with programmable phase shift. This is not to be confused with the programmable phase gate \cite{Micuda08,Mikova12} where only single-qubit rotation was programmed, while in our case we program a two-qubit operation by means of a third qubit. Tunable c-phase gate is a key ingredient for a number of protocols such as quantum routers \cite{Lemr13router,Vitelli13}, quantum state engineering \cite{Franson02klm,Lemr11klm} or controlled-unitary gates \cite{Lanyon09cpg,Lemr14cu}. By adding the programmability, we further develop this important gate, make it more versatile and therefore even a more powerful tool for quantum information processing. Conceptual scheme of the proposed gate is shown in Fig. \ref{fig:concept}. We adopt the following notation throughout the paper: $|\psi_T\rangle$, $|\psi_C\rangle$ and $|\psi_P\rangle$ denote the target, control and program qubits respectively. The program qubit takes the form of
\begin{equation}
\label{eq:program}
|\psi_P\rangle = \frac{1}{\sqrt{2}}\left(|0\rangle-\mathrm{e}^{i\varphi}|1\rangle\right),
\end{equation}
where $\varphi$ is the phase shift to be introduced by the gate if both the target and control qubits are in the logical state $|1\rangle$ as requested by the gate's definition [see Eq. (\ref{eq:cphase})]. The program qubit is destroyed by detection in the process while the target and control qubits leave the gate and can be used for further processing.

The paper is organized as follows: in Sec. \ref{Scheme} we derive the basic functioning of the gate and in Sec. \ref{Optimization} we show what techniques can be used to increase the success probability of the scheme. Subsequently in Sec. \ref{Feasibility} we discuss the scheme's experimental feasibility.

\section{Linear-optical scheme}
\label{Scheme}
\begin{figure}
\includegraphics[scale=1]{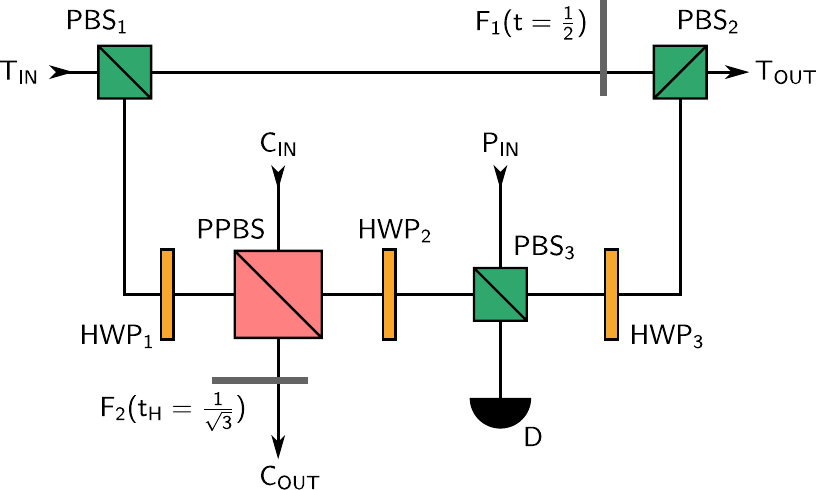}
\caption{\label{fig:setup} Linear-optical setup for the c-phase gate with programmable phase shift. The target, control and program qubit enter the setup at T$_\mathrm{IN}$, C$_\mathrm{IN}$ and P$_\mathrm{IN}$ respectively while the target and control output are denoted T$_\mathrm{OUT}$ and C$_\mathrm{OUT}$. The program qubit is detected by polarization sensitive detector D projecting it onto diagonally polarized state. Polarizing beam splitters PBS$_x$ ($x=1,2,3$) transmit horizontally polarized photons while reflecting vertical polarization. The partially polarizing beam splitter PPBS has unit transmissivity for horizontal polarization and $t_V=1/\sqrt{3}$ for vertical polarization (therefore the reflectivity for vertical polarization is $r_V=\sqrt{2/3}$). Filter F$_1$ is a neutral density filter with amplitude transmissivity $t_\mathrm{F1}=\frac{1}{2}$ while the filter F$_2$ only filters horizontal polarization with transmissivity $t_\mathrm{F2H}=1/\sqrt{3}$. Half-wave plates are rotated with respect to horizontal polarization by angles: HWP$_1$ @ $-9.2$ deg., HWP$_2$ and HWP$_3$ @ 22.5 deg. The gate succeeds if one photon leaves by the target output port, one photon by the control output port and one photon is detected by detector D.}
\end{figure}

Linear-optical scheme for a c-phase gate with programmable phase shift is depicted in Fig. \ref{fig:setup}. In this scheme we consider encoding logical states $|0\rangle$ and $|1\rangle$ into horizontal (H) and vertical (V) polarization states of individual photons. Similarly to the Lanyon {\it et al.} gate \cite{Lanyon09cpg}, we also introduce an auxiliary mode. In our case however, we use this mode for interaction between the target and program qubits. In this section, we derive the principle of operation by showing what transformation the gate implements on all four basis states as defined in Eq. (\ref{eq:cphase}) having simultaneously the phase shift $\varphi$ encoded in the state of the program qubit (\ref{eq:program}). The gate is necessarily probabilistic (all linear-optical c-phase gates are \cite{Kieling10cpg}) and its successful operation is heralded by observing one photon at each of the target and control output port and also by detecting a photon on detector D.
Let us start with the evaluation of the state $|00\psi_P\rangle$ (we maintain the order of qubits: target, control and program). The target photon impinges on the polarizing beam splitter PBS$_1$ that sends it to the upper path. There the target photon is subjected to a neutral density filter $F_1$ with amplitude transmissivity $t_{F1} = \frac{1}{2}$ and subsequently continues to the target output port by passing through the second polarizing beam splitter PBS$_2$. So far, one can write down the transformation of the state as
$$
|00\psi_P\rangle \rightarrow \frac{1}{2}|00\psi_P\rangle.
$$
Meanwhile the control photon is transmitted by the partially polarizing beam splitter PPBS (with transmissivity $t_{H}=1$ for horizontal polarization and $t_{V} = \sqrt{1-r_{V}^2} = \frac{1}{\sqrt{3}}$ for vertical polarization) and after being subjected to polarization filtering by the filter F$_2$ (filtering horizontal polarization with transmissivity $t_{F2{H}} = \frac{1}{\sqrt{3}}$ and letting the vertical polarization unfiltered $t_{F2{V}} = 1$) it leaves the setup by the control output port. At this point the transformation by the gate reads
$$
|00\psi_P\rangle \rightarrow \frac{1}{2\sqrt{3}}|00\psi_P\rangle.
$$
Finally the program photon impinges on the polarizing beam splitter PBS$_3$, where it gets transmitted and reflected with equal amplitude $1/\sqrt{2}$. Since the gate succeeds only if a photon is detected on detector D, only the transmission of the program photon through PBS$_3$ has to be taken into account. Considering the program photon is in the state (\ref{eq:program}), the overall state gets transformed into
$$
|00\psi_P\rangle \rightarrow \frac{1}{2\sqrt{6}}|000\rangle.
$$
Once the program photon leaves PBS$_3$, we project it onto diagonal polarization $|D\rangle = \frac{1}{\sqrt{2}}\left(|0\rangle+|1\rangle\right)$ resulting in the final form of the transformation
$$
|00\psi_P\rangle \rightarrow \frac{1}{4\sqrt{3}}|00D\rangle.
$$
This projection is needed to erase the which-path information about the photon detected on detector D.

In the same way, we now evaluate the transformation of the second state $|01\psi_P\rangle$. The only difference this time is in the control qubit. It impinges on the PPBS having vertical polarization and is therefore transmitted with amplitude $\frac{1}{\sqrt{3}}$ and reflected with amplitude $\sqrt{\frac{2}{3}}$. Only the transmission of the control photon by the PPBS contributes to the successful operation of the gate. Further to that, no attenuation of the vertically polarized control photon takes place on F$_2$. Having the same transformation for the target and program qubits as in the previous paragraph, one can identify the overall action of the gate
$$
|01\psi_P\rangle \rightarrow \frac{1}{4\sqrt{3}}|01D\rangle.
$$

A different situation happens for the third state $|10\psi_P\rangle$. The target photon is reflected by PBS$_1$ entering the lower path, where it is subjected to a half-wave plate HWP$_1$ oriented by $-9.2$ deg with respect to horizontal polarization. This wave plate transform the target photon in the following way
\begin{equation}
\label{eq:HWP1}
|1\rangle \rightarrow \frac{1}{2} |0\rangle + \frac{\sqrt{3}}{2}|1\rangle
\end{equation}
and thus causes the overall state to get transformed into
$$
|10\psi_P\rangle \rightarrow \frac{1}{2}|00\psi_P\rangle+ \frac{\sqrt{3}}{2}|10\psi_P\rangle.
$$
At this point the target and control photons interacts on the PPBS. Since the control photon is transmitted through the PPBS (having horizontal polarization in this case), we only take into account the transmission of the target photon to assure successful outcome of the gate
$$
|10\psi_P\rangle \rightarrow \frac{1}{2\sqrt{3}}|00\psi_P\rangle+ \frac{1}{2\sqrt{3}}|10\psi_P\rangle,
$$
where we have already incorporated the action of the polarization sensitive filter F$_2$. Now the target state enters a Hadamard transform implemented by a half-wave plate HWP$_2$ rotated by 22.5 degrees with respect to horizontal polarization providing target photon transformation of the form of
\begin{eqnarray}
\label{eq:Hadamard}
|0\rangle &\rightarrow& \frac{1}{\sqrt{2}}\left(|0\rangle+|1\rangle\right)\\\nonumber
|1\rangle &\rightarrow& \frac{1}{\sqrt{2}}\left(|0\rangle-|1\rangle\right),
\end{eqnarray}
which then translates into the overall state evolution
$$
|10\psi_P\rangle \rightarrow \frac{1}{\sqrt{6}}|00\psi_P\rangle.
$$
The target photon passes through PBS$_3$ having horizontal polarization. Therefore identically to the cases derived above, the gate can only succeed if the program photon passes through the PBS$_3$ and then gets projected onto diagonal polarization. Thus we obtain the transformation in the form of
$$
|10\psi_P\rangle \rightarrow \frac{1}{2\sqrt{6}}|00D\rangle.
$$
Finally, the target photon is again subjected to a Hadamard transform (HWP$_3$) resulting in
$$
|10\psi_P\rangle \rightarrow \frac{1}{4\sqrt{3}}\left(|00D\rangle+|10D\rangle\right).
$$
Only the target photon reflected by the PBS$_2$ leaves the gate by designated output port and thus we obtain the final form of the transformation
$$
|10\psi_P\rangle \rightarrow \frac{1}{4\sqrt{3}}|10D\rangle.
$$

To complete our analysis, we now evaluate the transformation of the last basis state $|11\psi_P\rangle$. Similarly to the previous case, the target photon gets reflected on PBS$_1$ and transformed by the half-wave plate HWP$_1$ according to prescription (\ref{eq:HWP1}). The overall state thus takes the form of
$$
|11\psi_P\rangle \rightarrow \frac{1}{2}|01\psi_P\rangle+ \frac{\sqrt{3}}{2}|11\psi_P\rangle.
$$
At this point two-photon interference on the PPBS takes place resulting in
\begin{eqnarray}
|11\psi_P\rangle &\rightarrow& \frac{1}{2\sqrt{3}}\left( |01\psi_P\rangle+ |11\psi_P\rangle - 2|11\psi_P\rangle\right)\nonumber\\
&=& \frac{1}{2\sqrt{3}}\left(|01\psi_P\rangle- |11\psi_P\rangle\right),
\end{eqnarray}
where only the terms contributing to success of the gate are shown. Note that when both the target and control photons enter the PPBS in vertical polarization state, the interference of both the photons being transmitted and both the photons being reflected (Hong-Ou-Mandel interference) occurs introducing a phase shift $\pi$ to the term $|11\psi_P\rangle$ \cite{Kiesel05}. By means of the subsequent Hadamard transform in the target mode (HWP$_2$), the state transforms into
$$
|11\psi_P\rangle \rightarrow \frac{1}{\sqrt{6}}|11\psi_P\rangle.
$$
On PBS$_3$ the target photon gets reflected (being vertically polarized) and so only the program photon reflection can contribute to the gate's successful operation. This means that just its vertical polarization term contributes yielding the overall state in the form of
$$
|11\psi_P\rangle \rightarrow -\mathrm{e}^{i\varphi}\frac{1}{2\sqrt{3}}|111\rangle
$$
which after projecting the photon in program mode onto diagonal polarization gives
$$
|11\psi_P\rangle \rightarrow -\mathrm{e}^{i\varphi}\frac{1}{2\sqrt{6}}|11D\rangle.
$$
Action of the Hadamard gate in the target mode (HWP$_3$) and reflection of the target photon on PBS$_2$ to its output port results in the final transformation
$$
|11\psi_P\rangle \rightarrow \mathrm{e}^{i\varphi}\frac{1}{4\sqrt{3}}|11D\rangle.
$$
In contrast to the three previous cases, the state is now phase-shifted by angle $\varphi$ exactly as prescribed in (\ref{eq:cphase}).

We have shown that the setup depicted in Fig.~\ref{fig:setup} implements the tunable c-phase gate with the phase shift programmed by the program qubit. This has been demonstrated on all four basis states of the control and target qubits together with an arbitrary program state. Since all transformations are linear, the entire operation holds also for any superposition of the above mentioned four basis states. The success probability of the gate is $1/(4\sqrt{3})^2=\frac{1}{48}$ and is state independent (in order not to deform superpositions of basis states). In the next section, we will consider potential improvements to the setup in order to increase the success probability. 


\section{Increasing the success probability}
\label{Optimization}
\begin{figure}
\includegraphics[scale=1]{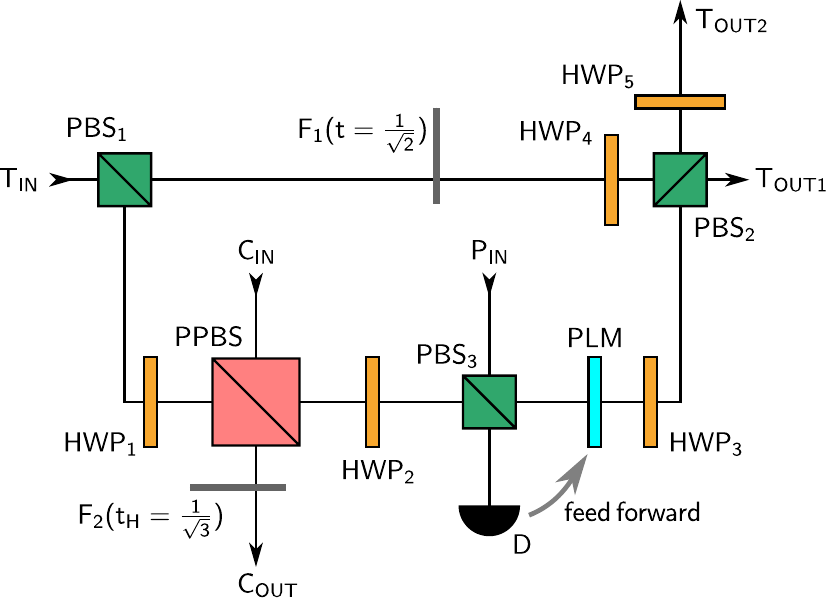}
\caption{\label{fig:setup_opti} Optimized setup for the programmable c-phase gate. The components are designated as in Fig. \ref{fig:setup} with the newly added half-wave plates HWP$_4$ (rotated by 22.5 deg.) and HWP$_5$ (rotated by 45 deg.) and a phase modulator PLM implementing a feed forward operation.}
\end{figure}
So far we have discussed the basic scheme for the programmable c-phase gate which is experimentally also the most easily implementable. Its success probability is however more then four times lower then the success probability of the optimal non-programmable c-phase gate. In this section we will discuss two optimization approaches allowing for considerable improvement in success probability (see optimized scheme in Fig. \ref{fig:setup_opti}).

Firstly, one can increase the success probability of the program photon projection implemented before its detection. As derived in the previous section, the gate succeeds if the program photon is projected onto diagonal polarization and detected by D. This way, we neglect half of the cases corresponding to the program photon being projected onto anti-diagonal polarization [$|A\rangle = \frac{1}{\sqrt{2}}\left(|0\rangle-|1\rangle\right)$]. As experimentally demonstrated on a simpler unconditional programmable gate \cite{Mikova12}, one can increase the success probability by a factor of two if the anti-diagonal projections of the program photon are included. In such cases a feed-forward transformation $|1\rangle\rightarrow-|1\rangle$ has to by applied to the target photon immediately as it exits PBS$_3$ \cite{Bula13}. Such transformation can be achieved using for instance a phase modulator (PLM) \cite{Mikova12}.

The second possibility to increase the overall success probability again by a factor of two is to use both the output ports of PBS$_2$ (designated T$_\mathrm{OUT1}$ and T$_\mathrm{OUT2}$ in Fig. \ref{fig:setup_opti}). In this case, the amplitude transmissivity of the filter F$_1$ shall be reduced to $t_\mathrm{F1}=\frac{1}{\sqrt{2}}$ and a half-wave plate HWP$_4$ inserted behind it. This newly added wave-plate is rotated by 22.5 degrees with respect to horizontal polarization to implement the Hadamard transform (\ref{eq:Hadamard}). The target state at T$_\mathrm{OUT1}$ is thus kept unchanged, but it allows for the target photon to leave also by the output port T$_\mathrm{OUT2}$. The target photon exiting PBS$_2$ by its second output is however polarization-swapped with respect to the target photon in the first output. A half-wave plate HWP$_5$ (rotated by 45 degrees) is therefore inserted to the output port T$_\mathrm{OUT2}$ to perform the swap operation ($|0\rangle\rightarrow|1\rangle$, $|1\rangle\rightarrow|0\rangle$). The second output port can be used only if one does not require the target photon to leave by a specified output. Such situation occurs for instance if the target qubit is immediately measured after being processed by the gate (the gate is the last element in a quantum circuit). The measurement apparatus can then be installed to both output ports.

If only one of the mentioned optimization strategies is used, the success probability of the gate increases to 1/24 and if both of them are used it reaches the value of 1/12 (for all phases $\varphi$). Note that optimal non-programmable c-phase gate has the minimum success probability being about 1/11 for $\varphi$ close to $\frac{2\pi}{3}$ \cite{Lemr11cpg}. If completely optimized, the programmable gate thus performs with almost the same probability as the optimal non-programmable gate at its success probability minimum.

\section{Experimental feasibility}
\label{Feasibility}
In this section, we discuss the feasibility of the proposed scheme based on the current level of technological development in linear-optical quantum information processing with discrete photons. Firstly, in order to achieve any linear-optical quantum gate, one needs to generate adequate input photon state. Our gate requires three photons each bearing one polarization-encoded qubit. Generation of three separate photons has already been achieved in various experiments. Either one photon pair from spontaneous parametric down-conversion (SPDC) is combined with one additional photon from attenuated fundamental laser beam \cite{Slodicka2009} or two photon pairs are generated via SPDC with one photon serving just as a trigger \cite{Vitelli13}. With either of these techniques one can generate input states with sufficient fidelity, typically more than 95\%, with repetition rate at about 1 per 5\,s \cite{Dobek2013}.

In the next step, we asses the feasibility of stabilizing the proposed scheme. There are two types of stabilizations required: two-photon temporal and spatial overlap stabilization (Hong-Ou Madel interference \cite{HOM}) and single-photon interference stabilization. The first mentioned has to be achieved with precision of about 1/50 of the photons wave packet length. Such wave packet is typically \SI{100}{\micro\meter} long (FWHM) in space and therefore stability with precision in units of \si{\micro\meter} suffices. One can use motorized translation to achieve this task. From our experience, the two-photon overlap is stable for about one hour \cite{Lemr11cpg}. After that, one needs to perform two-photon interference measurement and observe the Hong-Ou-Mandel dip to reset the actual position of maximal two-photon overlap. Since two-fold coincidences occur much more frequently then three-fold or four-fold ones, this restabilization measurement, when optimized, can take no more that a minute. In our scheme, we require to stabilize the two-photon overlap on the PPBS and on PBS$_3$. Note that the stabilization procedures can be performed separately and consecutively and hence the need for two of them does not pose significant technological difficulties. There is also one instance of single-photon interference occurring in the setup between the polarizing beam splitters PBS$_1$ and PBS$_2$. In contrast to the two-photon overlap, the single-photon interference needs to be stabilized typically to at least $\lambda/50$, where $\lambda$ is the photon's wavelength. Such precision requires combining both motorized translation for larger steps with piezo translation for fine adjustments. In a typical bulk interferometer on decimeter scale, the single-photon stability only lasts for less than a minute \cite{Lemr11cpg}. One can however significantly increase such time by replacing classical interferometer by a compact design using beam dividers \cite{Micuda13toffoli}. In general, there are two ways of how to perform stabilization of this kind. One option is to use the individual photons themselves and perform a set of measurements for various settings of the piezo shift. Note that single-photon detections occur even with much higher rate that two-fold coincident detections and therefore there is usually fair amount of signal to work on \cite{Lemr11cpg,Lemr10klm}. Other experimentalists prefer using a strong laser beam propagating along the weak quantum signal. This strong signal is at sufficiently distant wavelength to allow mixing and subsequent decoupling from the quantum signal using dichroic filters. In this case the stabilization can be performed ``on-line'' during the entire experiment \cite{Xavier2011interference}. Using either of the above described techniques, it is quite feasible to stabilize the single-photon interference in the proposed setup.

Finally, the feasibility considerations have to be dedicated to the final detection procedure. In order to be experimentally implementable, the detection has to be robust against non-unit quantum detection efficiency of typical detectors and technological losses (e.g. back-reflection, coupling efficiency). This requirement rules out vacuum detection-based schemes (schemes where success is heralded by vacuum detection or no-detection) \cite{VanMeter07}. Similarly, photon-number resolving detection is not completely reliable because of detection (in)efficiency \cite{Rehacek03loop} and technological losses. In our case however, the scheme only requires post-selection on three-fold coincidence detections and thus the quantum efficiency of the detectors only affects the detection rate and not the detected quantum state. This is a key feature of the proposed scheme with respect to its feasibility.

\section{Conclusions}
\label{Conclusions}
In this paper, we have provided a linear-optical scheme for a programmable c-phase gate. The phase shift introduced by this gate is set by the state of a program qubit which makes the gate more versatile than previously implemented tunable c-phase gates with phase shift hard-coded to the setup setting. The setup is designed with experimental feasibility in mind. It does not require photon-number resolving detectors nor post-selection on vacuum detection and is therefore implementable with current level of know-how in experimental linear-optical quantum information processing.

We have also presented two optimization options. Each of them allows doubling the overall success probability of the gate. Using both these optimizations, the gate succeeds with probability of 1/12 which is close to the success probability of a non-programmable tunable c-phase gate \cite{Lemr11cpg}. Further the two optimization steps can be used independently allowing to obtain success probability of 1/24 if only one of them is used. The first optimization method consists of applying an experimentally feasible feed-forward operation. The second optimization involves using both output ports of the final polarizing beam splitter in the target mode.


\section*{Acknowledgment}
\label{Acknowledgment}
The authors thank Jára Cimrman for his helpful suggestions.
K.~L. acknowledges support by the Czech Science Foundation (Grant No.
13-31000P). K.~B. acknowledges support by the Foundation for Polish Science and
the Polish National Science Centre under grant No.
DEC-2011/03/B/ST2/01903. Finally, the authors acknowledge the project No. LO1305 of the
Ministry of Education, Youth and Sports of the Czech Republic.


\begin{thebibliography}{00}

\bibitem{Nielsen_QCQI}
M.~A.~Nielsen and I.~L.~Chuang, 
Quantum Computation and Quantum Information,
Cambridge University Press, Cambridge, 2002.

\bibitem{Zeilinger_QIP}
D.~Bouwmeester, A.~Ekert, A.~Zeilinger,
The Physics of Quantum Information,
Springer, Heidelberg, 2001.

\bibitem{Barenco95universal}
A.~Barenco, C.~H.~Bennett, R.~Cleve, D.~P.~DiVincenzo, N.~Margolus, P.~Shor, 
T.~Sleator, J.~A.~Smolin, and H.~Weinfurter,
Elementary gates for quantum computation, 
Phys.~Rev.~A {\bf 52}, 3457 (1995).

\bibitem{Shende04optim}
V.~V.~Shende, I.~L.~Markov, and S.~S.~Bullock,
Minimal universal two-qubit controlled-NOT-based circuits, 
Phys.~Rev.~A {\bf 69}, 062321 (2004).

\bibitem{Cory97}
D.~Cory, M.~Price, A.~Fahmy, and T.~Havel,
Ensemble quantum computing by NMR spectroscopy, 
Proc.~Natl.~Acad.~Sci.~U.S.A. {\bf 94}, 199 (1997).


\bibitem{Schmidt-Kaler03}
F.~Schmidt-Kaler, H.~Haffner, M.~Riebe, S.~Gulde, G.~P.~T.~Lancaster, 
T.~Deuschle, C.~Becher, C.~F.~Roos, J.~Eschner, and R.~Blatt, 
Realization of the Cirac-Zoller controlled NOT quantum gate, 
Nature (London) {\bf 422}, 408 (2003). 

\bibitem{Plantenberg07}
J.~H.~Plantenberg, P.~C.~de~Groot, C.~J.~P.~M.~Harmans, and J.~E.~Mooij,
Demonstration of controlled-NOT quantum gates on a pair of superconducting quantum bits,
Nature (London) {\bf 447}, 836 (2007). 


\bibitem{Ralph02}
T.~C.~Ralph, N.~K.~Langford, T.~B.~Bell, and A.~G.~White,
Linear optical controlled-NOT gate in the coincidence basis,
Phys.~Rev.~A {\bf 65}, 062324 (2002).

\bibitem{OBrien03}
J.~L.~O'Brien, G.~J.~Pryde, A.~G.~White, T.~C.~Ralph, and D.~Branning,
Demonstration of an all-optical quantum controlled-NOT gate,
Nature (London) {\bf 426}, 264 (2003).

\bibitem{Kiesel05}
N.~Kiesel, C.~Schmid, U.~Weber, R.~Ursin, and H.~Weinfurter,
Linear Optics Controlled-Phase Gate Made Simple, 
Phys.~Rev.~Lett. {\bf 95}, 210505 (2005).

\bibitem{Kok07}
P.~Kok, W.~J.~Munro, K.~Nemoto, T.~C.~Ralph, J.~P.~Dowling, and G.~J.~Milburn,
Linear optical quantum computing with photonic qubits, 
Rev. Mod. Phys. {\bf 79}, 135 (2007). 

\bibitem{Bartkowiak10}
M.~Bartkowiak, and A.~Miranowicz,
Linear-optical implementations of the iSWAP and controlled NOT gates based on conventional detectors, 
J.~Opt.~Soc.~Am.~B {\bf 27}, 2369 (2010).

\bibitem{Lanyon09cpg}
B.~P.~Lanyon, M.~Barbieri, M.~P.~Almeida, T.~Jennewein, T.~C.~Ralph, K.~J.~Resch, G.~J.~Pryde, J.~L.~O'Brien, A.~Gilchrist, and A.~G.~White, 
Simplifying quantum logic using higher-dimensional Hilbert spaces,
Nat. Phys. {\bf 5}, 134 (2009).

\bibitem{Kieling10cpg}
K.~Kieling, J.~L.~O'Brien, J.~Eisert,
On photonic controlled phase gates, 
New~J.~Phys. {\bf 12}, 013003 (2010).

\bibitem{Lemr11cpg}
K.~Lemr, A.~Černoch, J.~Soubusta, K.~Kieling, J.~Eisert, and M.~Dušek,
Experimental implementation of the optimal linear-optical controlled phase gate, 
Phys.~Rev.~Lett. {\bf 106}, 013602 (2011).

\bibitem{proggates}
M.~A.~Nielsen and I.~L.~Chuang, Programmable Quantum Gate Arrays, Phys.~Rev.~Lett. {\bf 79}, 321 (1997);
G.~Vidal, L.~Masanes, and J.~I.~Cirac, Storing Quantum Dynamics in Quantum States: A Stochastic Programmable Gate, Phys.~Rev.~Lett. {\bf 88}, 047905 (2002);
M.~Hillery, V.~Bužek, and M.~Ziman, Probabilistic implementation of universal quantum processors, Phys.~Rev.~A {\bf 65}, 022301 (2002).

\bibitem{Micuda08}
M.~Mičuda, M.~Ježek, M.~Dušek, and J.~Fiurášek,
Experimental realization of a programmable quantum gate,
Phys.~Rev.~A {\bf 78}, 062311 (2008).

\bibitem{Mikova12}
M.~Miková, H.~Fikerová, I.~Straka, M.~Mičuda, J.~Fiurášek, M.~Ježek, M.~Dušek,
Increasing efficiency of a linear-optical quantum gate using electronic feed-forward,
Phys.~Rev.~A {\bf 85}, 012305 (2012).


\bibitem{Lemr13router}
K.~Lemr, K.~Bartkiewicz, A.~Černoch, and J.~Soubusta,
Resource-efficient linear-optical quantum router,
Phys.~Rev.~A {\bf 87}, 062333 (2013).

\bibitem{Vitelli13}
Ch.~Vitelli, N.~Spagnolo, L.~Aparo, F.~Sciarrino, E.~Santamato, and L.~Marrucci,
Joining the quantum state of two photons into one,
Nat. Phot. {\bf 7}, 521 (2013).

\bibitem{Franson02klm}
J.~D.~Franson, M.~M.~Donegan, and B.~C.~Jacobs,
Generation of entangled ancilla states for use in linear optics quantum computing,
Phys.~Rev.~A {\bf 69}, 052328 (2004).

\bibitem{Lemr11klm}
K.~Lemr,
Preparation of Knill-Laflamme-Milburn states using tunable controlled phase gate,
J. Phys. B: At. Mol. Opt. Phys. {\bf 44}, 195501 (2011).

\bibitem{Lemr14cu}
K.~Lemr, K.~Bartkiewicz, A.~Černoch, M.~Dušek, and J.~Soubusta,
Experimental implementation of optimal linear-optical controlled-unitary gates,
Phys.~Rev.~Lett. {\bf 114}, 153602 (2015).

\bibitem{Bula13}
M.~Bula, K.~Bartkiewicz, A.~Černoch, and K.~Lemr, 
Entanglement-assisted scheme for nondemolition detection of the presence of a single photon, 
Phys.~Rev.~A {\bf 87}, 033826 (2013). 

\bibitem{Slodicka2009}
L.~Slodička, M.~Ježek, and J.~Fiurášek,
Experimental demonstration of a teleportation-based programmable quantum gate,
Phys.~Rev.~A {\bf 79}, 050304(R) (2009).

\bibitem{Dobek2013}
K.~Dobek, M.~Karpinski, R.~Demkowicz-Dobrzanski, K.~Banaszek, P.~Horodecki,
Experimental generation of complex noisy photonic entanglement,
Las.~Phys. {\bf 23}, 025204 (2013).

\bibitem{HOM}
C.~K.~Hong, Z.~Y.~Ou, and L.~Mandel,
Measurement of subpicosecond time intervals between two photons by interference,
Phys.~Rev.~Lett. {\bf 59}, 2044 (1987).

\bibitem{Micuda13toffoli}
M.~Mičuda, M.~Sedlák, I.~Straka, M.~Miková, M.~Dušek, M.~Ježek, and J. Fiurášek,
Efficient Experimental Estimation of Fidelity of Linear Optical Quantum Toffoli Gate, 
Phys.~Rev.~Lett.~{\bf 111}, 160407 (2013).

\bibitem{Lemr10klm}
K.~Lemr, A.~Černoch, J.~Soubusta, and J.~Fiurášek,
Experimental preparation of two-photon Knill-Laflamme-Milburn states,
Phys.~Rev.~A {\bf 81}, 012321 (2010).

\bibitem{Xavier2011interference}
G.~B.~Xavier, J.~P.~von~der~Weid,
Stable single-photon interference in a 1 km fiber-optical Mach-Zehnder interferometer with continuous phase adjustment,
Opt.~Lett. {\bf 36}, 1764 (2011).

\bibitem{VanMeter07}
N.~M.~VanMeter, P.~Lougovski, D.~B.~Uskov, K.~Kieling, J.~Eisert, J.~P.~Dowling,
General linear-optical quantum state generation scheme: Applications to maximally path-entangled states,
Phys.~Rev.~A {\bf 76}, 063808 (2007).

\bibitem{Rehacek03loop}
J.~Řeháček, Z.~Hradil, O.~Haderka, J.~Peřina Jr., M.~Hamar,
Multiple-photon resolving fiber-loop detector,
Phys.~Rev.~A {\bf 67}, 061801(R) (2003).





\end{thebibliography}
\end{document}